\documentclass[twocolumn,aps,a4paper,superscriptaddress,showkeys,reprint]{revtex4-1}
\usepackage{graphicx}
\usepackage{latexsym}   
\usepackage{amsmath,amssymb,amsthm}

\setcitestyle{super}

\begin{document}

\title{Influence of network topology on the swelling of polyelectrolyte nanogels}

\author{\firstname{L.} G. \surname{Rizzi}} %
\email{lerizzi@ufv.br}
\affiliation{Departamento de F\'isica, Universidade Federal de Vi\c{c}osa, 36570-900, Vi\c{c}osa, MG, Brazil.}
\affiliation{Instituto de F\'isica, Universidade Federal do Rio Grande do Sul, CP 15051, 91501-970, Porto Alegre, RS, Brazil.}
\author{\firstname{Y.} \surname{Levin}} %
\email{levin@if.ufrgs.br}
\affiliation{Instituto de F\'isica, Universidade Federal do Rio Grande do Sul, CP 15051, 91501-970, Porto Alegre, RS, Brazil.}

\date{\today}


\begin{abstract}
	It is well-known that the swelling behavior of ionic nanogels depends on their cross-link density, however it is unclear how different topologies should affect the response of the polyelectrolyte network.~Here we perform Monte Carlo simulations to obtain the equilibrium properties of ionic nanogels as a function of salt concentration $C_s$ and the fraction $f$ of ionizable groups in a polyelectrolyte network formed by cross-links of functionality $z$.
	Our results indicate that the network with cross-links of low connectivity result in nanogel particles with higher swelling ratios.
	We also confirm a de-swelling effect of salt on nanogel particles.
\end{abstract}



\maketitle


\section{INTRODUCTION}

	Ionic microgel and nanogel particles formed by cross-linked polyelectrolyte networks displays many remarkable properties 
	which make them suitable for applications in drug-delivery systems, where molecules can be encapsulated and then released at specific targets~\cite{oh2008progrep,chacko2012rev}.
	This is possible through a swelling (or de-swelling) process, where solvent molecules flow into (or leave) the cross-linked network.
	Experimentally~\cite{microgel2011book,riest2012zphyschem} it is well-known that the properties of ionic microgels and nanogels particles can be significantly affected by the changes in temperature, solvent quality, salt concentration, ionic strength, and degree of cross-linking.
	Although these effects have been extensively studied in ionic macrogels~\cite{escobedo1999physrev,mann2005jcp,yin2008jcp,yin2009jcp,quesada2012macromol}, little is known about ionic nanogels formed by finite-size cross-linked polyelectrolyte networks~\cite{quesadaperez2011softmatt,quesadaperez2014jcp}.

	Theoretically, the swelling processes in polyelectrolyte networks have been studied~\cite{likos2001physrep,quesadaperez2011softmatt,riest2012zphyschem} using Flory's theory~\cite{flory1953book},
	which combines the mixing term, electrostatic, and elastic contributions into the total free energy of the system.
	An important aspect of the theory is the relationship between the (effective) number of chains of the network $N_{\text{eff}}$ and the elastic free energy~\cite{flory1985britpolj}, {\it i.e.}~$\Delta F_{\text{elastic}} \propto N_{\text{eff}}$.
	Clearly, the unambiguous interpretation of this relationship in terms of the network topology and its validation is relevant to the development of predictive theories, {\it e.g.}~\cite{levin2002pre,colla2014jcp}.
	In principle, such relationship can be verified by computer simulations considering explicit network structures, but to our knowledge current studies have so far only exploited polyelectrolyte networks with diamond-like structures ({\it i.e.}~with tetrafunctional cross-links)~\cite{claudio2009jcp,quesadaperez2013softmatter,kobayashi2014polymers,quesadaperez2014jcp,adroher2015macromol}.

	Here we explore this issue by considering three types of nanogel particles generated by polyelectrolyte networks of different topologies characterized by the functionality (also known as coordination number or connectivity) $z$ of the cross-links.
	In particular, we will investigate how the equilibrium properties of the nanosized gel particles are affected by the fraction $f$ of the ionizable groups and salt concentration $C_s$ in solution. 
	These important effects have also been explored by experiments\cite{english1996jcp,fan2010jcolloid,nerapusri2006lang,nisato1999langmuir,horkay2000biomacromol,xia2003macromol,dubrovskii1997macromol,lopezleon2006jphyschemB,capriles2008jphyschemB} and theory~\cite{pincus1991macromol,rydzewski1990contin,fernandez2001jcp,levin2002pre,fernandez2003jcp,victorov2006pccp,quesadaperez2011softmatt,sing2013macromol,yigit2012langmuir,colla2014jcp}.


\section{MODEL AND METHODS}

\begin{figure}[!t]
\centering
\includegraphics[scale=0.39,bb=0 0 550 360]{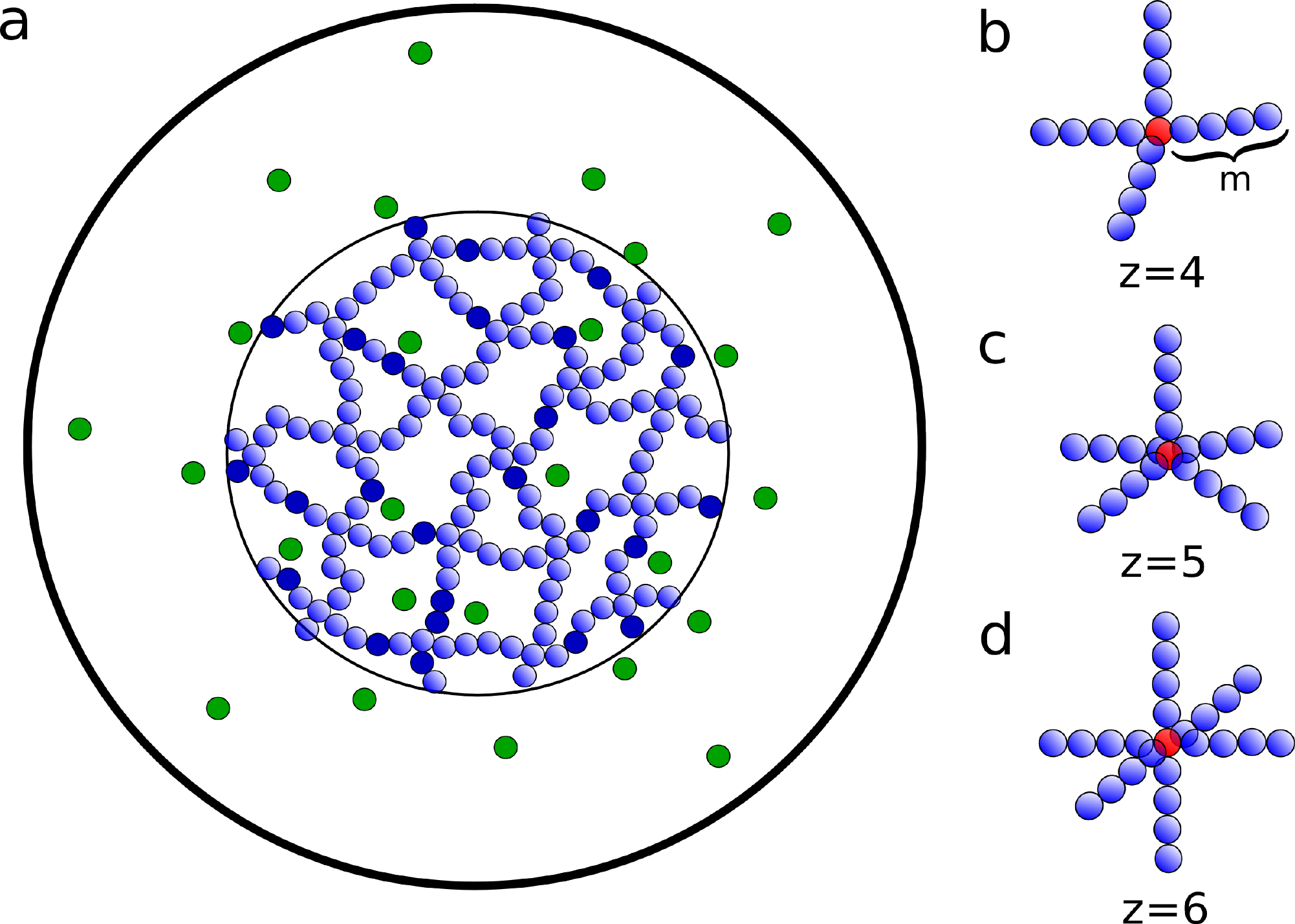}
\caption{(a) Nanogel particle is defined by a polyelectrolyte network inside a WS cell (outer circle) with radius $R$. The dissociation fraction $f$ determines the number of charged monomers with ionizable (anionic) groups (shown in dark blue), and the number $Z=fN_{\text{mon}}$ counterions in solution (shown in green); neutral monomers are displayed as light blue spheres. Inner circle represents the radius of nanogel particle defined as the radius of gyration $R_g$ of the polyelectrolyte network. (b-d) Schematic representation of cross-links with functionality $z$ that connect chains with $m$ monomers and are used to generate the network with different topologies.}
\label{systemdef:zcoord}
\end{figure}

	As illustrated in Fig.~\ref{systemdef:zcoord}(a), the nanogel particle is modeled as a polyelectrolyte network with $N_{\text{mon}}$ monomers inside a spherical Wigner-Seitz (WS) cell of volume $V=1/\rho$, where $\rho$ is the number density of nanogel particles in  solution.
	The radius of the WS cell is then  $R = (3/4 \pi \rho)^{1/3}$.
	Some monomers of the network are ionizable (anionic) and will dissociate releasing a counterion into solution.
	The fraction of dissociation $f$ determines the number of negatively charged monomers $Z = f N_{\text{mon}}$ in the network, and charge neutrality requires $Z$ counterions in solution.
	The counterions are allowed to diffuse everywhere inside the WS cell.  For simplicity, we will assume that all the ions  are spherical while the solvent, water, is modeled as a dielectric continuum.  

	In order to investigate the effect of topology on the swelling of the nanogel particle, we study three different networks that are arbitrarly generated from a regular structures determined by the functionality $z$ of the cross-links.
	As illustrated in Fig.~\ref{systemdef:zcoord}(b)-(d), the networks are built considering cross-links with $z=4$, $5$, and $6$, that are connected to each other through chains with $m$ monomers.
	The final networks are obtained by cropping the large regular template structures (with the number of monomers larger than $N_{\text{mon}}$), so that only the $N_{\text{mon}}$ monomers inside a spherical volume comprise the nanogel particle.
	Importantly, we have selected such spherical volume in a way that all the different networks are formed by approximately the same number of monomers $N_{\text{mon}}$, which leads to different number of cross-links $N_c$ with functionality $z$, as shown in Table~\ref{parameters}.
	Note that this procedure leads to networks with higher effective cross-link line density $\rho_c =N_c/N_{\text{mon}}$ as $z$ is lowered.
	Also, the cropping procedure inevitably results in $N_d$ dangling chains (which could have less than $m$ monomers) near the surface of the polyelectrolyte network, as illustrated in Fig.~\ref{systemdef:zcoord}(a).

\begin{table}[!b]
\caption{Parameters of the generated networks used in the simulations; $z$: functionality, $N_{\text{mon}}$: number of monomers, $N_c$: number of cross-links that connect $z$ chains, $N_d$: number of dangling ends, $\rho_c$: effective cross-link line density.}
 \begin{tabular}{ccccc}
  \hline\\[-0.37cm]
  \hline
~~ $z$ ~~ & ~~~ $N_{\text{mon}}$ ~~~ &  ~~~ $N_c$ ~~~ & ~~~ $N_d$ ~~~ & ~~ $\rho_{c}=N_c/N_{\text{mon}}$ ~~ \\ \hline
  4  &      2808        &   152  &  104   &   0.054  \\
  5  &      2524        &   108  &  108   &   0.043  \\
  6  &      2715        &    93  &   78   &   0.035  \\
 \hline\\[-0.37cm]
  \hline
\end{tabular}
\label{parameters}
\end{table}

	Next we introduce the interaction potentials between the components of the system.
	To describe a nanogel particle immersed in an implicit athermal solvent, 
where there is a slight preference of polyelectrolytes to be surrounded by solvent molecules,
our model assumes that all monomers in the network interact via a non-bonded, shifted and truncated repulsive Lennard-Jones potential, {\it i.e.} the Weeks-Chandler-Andersen~\cite{wca1971jcp} (WCA)
potential commonly used in simulations of polyelectrolytes (see {\it e.g.}~\cite{stevens1995jcp,santos2013softmat}), which is given by
\begin{equation}
\beta U_{\text{LJ}}(r_{ij}) = 4 \varepsilon \left[ \left(\frac{\sigma}{r_{ij}}\right)^{12} -  \left(\frac{\sigma}{r_{ij}}\right)^{6} + \frac{1}{4} \right]~,
\end{equation}
if the distance $r_{ij}$ between $i$ and $j$ monomers is less than a cutoff radius $r_c=2^{1/6}\sigma$, or zero otherwise; here $\varepsilon$ and $\sigma$ are parameters that determine the energy and length scales, respectively; and $\beta=1/k_BT$, where $T$ is the temperature and $k_B$ is the Boltzmann's constant.
	In addition, adjacent monomers in the network (which are defined {\it a priori} by construction) interact via a finitely extensible nonlinear elastic (FENE) potential~\cite{stevens1995jcp,santos2013softmat},
\begin{equation}
 \beta U_{\text{FENE}}(r_{ij}) = - 0.5 k_s R_0^2 \ln\left[1 - \frac{\left(r_{ij}-r_0\right)^2}{R_0^2}\right]~~,
\label{fenepotential}
\end{equation}
where $k_s = 7 \varepsilon/ \sigma^2$ is the spring constant, $r_0$ defines the minimum of the potential ({\it i.e.} its equilibrium distance), and $R_0 = 2 \sigma$ is the maximum extension allowed.
	All charged particles ({\it i.e.}~monomers and ions) interact via an electrostatic potential which can be written as
\begin{equation}
\beta  U_{\text{Coulomb}}(r_{ij}) = \lambda_B \frac{\alpha_i \alpha_j}{r_{ij}} ~~,
\label{coulombpot}
\end{equation}
where $\lambda_B$ is the Bjerrum length and $\alpha_i = \pm 1$ or 0, depending of the charges of the interacting particles.
	We also include a hard core (excluded volume) potential between all particles ({\it i.e.}~monomer-monomer, ion-monomer and ion-ion) which is $+\infty$ if $r_{ij}<d_{\min}$ or 0, otherwise.
	This potential precludes any two particles to be closer than a distance $d_{\min}$.

	In order to obtain the equilibrium properties of the system we perform Monte Carlo (MC) simulations with standard Metropolis acceptance criteria~\cite{frenkelbook}, where the transition probability
is given by $p(\{\vec{r}_{ij}\}_{\text{old}}\,\rightarrow\,\{\vec{r}_{ij}\}_{\text{new}})=\min\left(1,\exp{-\beta [ U(\{\vec{r}_{ij}\}_{\text{new}}) - U(\{\vec{r}_{ij}\}_{\text{old}}) ]}\right)$, with $\beta U(\{\vec{r}_{ij}\}_{\text{new}})$ and $\beta U(\{\vec{r}_{ij}\}_{\text{old}})$ being the total energetic contribution ({\it i.e.}~WCA, FENE, electrostatic, and hard core potentials) evaluated for the new and old configurations of the system, respectively.
	New configurations are proposed by attempting to move a single particle and a MC step is defined after all particles in the system have attempted a move.
	Only diffusive-like random movements are considered, which means that a particle could move in any direction within a maximum displacement with modulus $d_{\max}$.
	The value of $d_{\max}=4$~\AA~is considered here since it provided the optimal choice from previous studies~\cite{santos2013softmat}.



\section{SIMULATION RESULTS}

	Next we present results for a salt-free solution, {\it i.e.}~when $C_s=0$ and the only dissociated ions in the system are those from the polyelectrolyte network.
	For all topologies consider, the networks are generated with chains including $m=8$ monomers between the cross-links.
	In all simulations the nanogel concentration is set at $\rho=10^{-8}~$\AA$^{-3}$ (approximately $17~\mu$M), and the parameters of the potentials are given by 
$\varepsilon=0.8333$, 
$\sigma=2r_{\text{Na}^{+}}=4$~\AA, 
$\lambda_B =7.2$, 
$r_0=4$~\AA,
and 
$d_{\min}=4$~\AA. 
	For each value of dissociation fraction $f$, the mean values and error bars of the quantities presented below were evaluated using configurations taken from 50 independent simulations (different seeds) with runs amounting $0.5 - 1.0 \times 10^6$ MC steps.
	To avoid correlations, we skip $10^4$ MC steps between samples in the same run, so that a measurement for a given fraction $f$ corresponds to at least 2500 different configurations.

\begin{figure}[!b]
\centering
\includegraphics[scale=0.24,bb=0 0 550 1100]{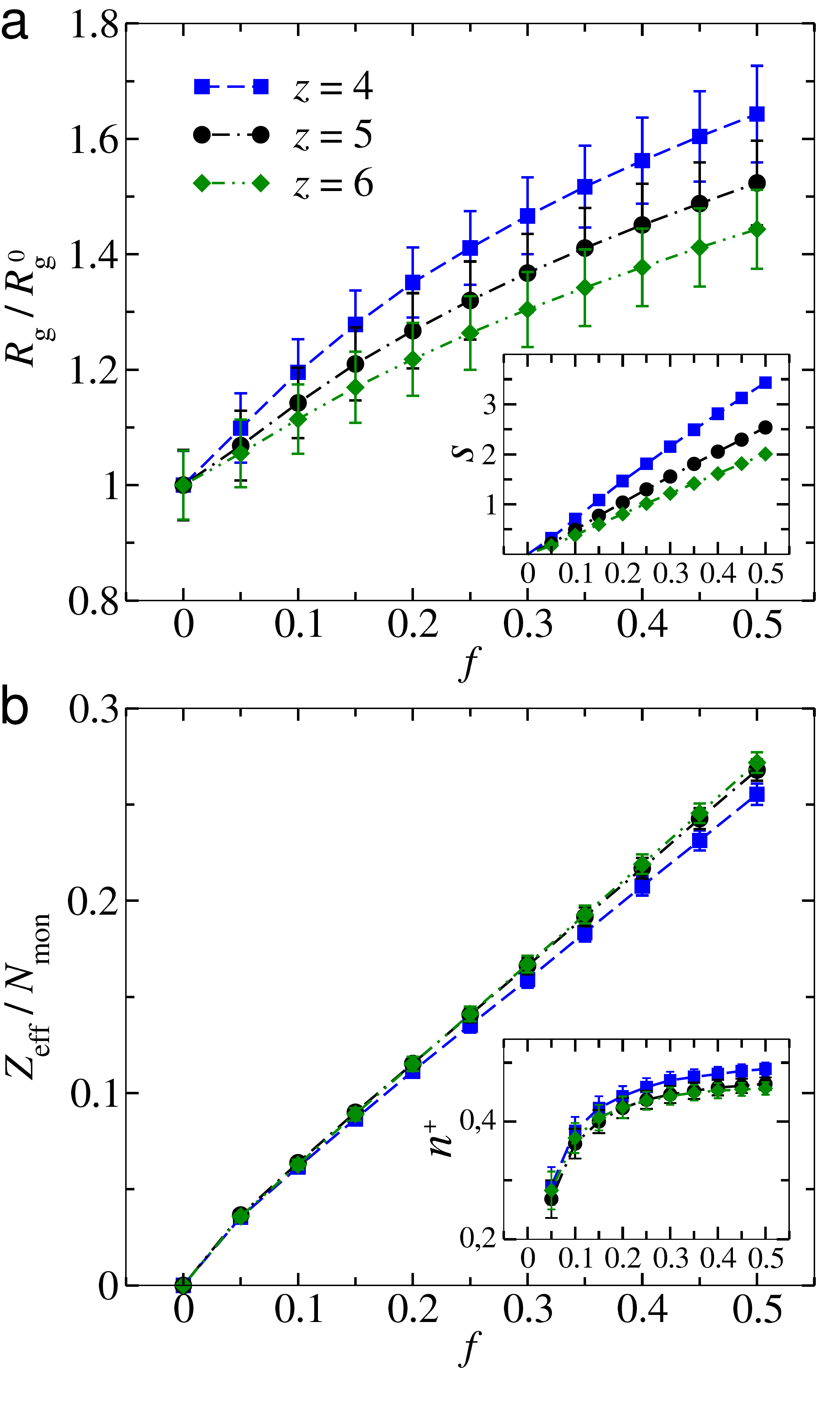}
\caption{Effect of the dissociation fraction $f$. (a) Relative nanogel radius $R_g/R_g^0$ (Inset: swelling ratio $S$); and (b) effective nanogel charge $Z_{\text{eff}}$ (Inset: relative number of positive charges inside a nanogel particle $n^{+}=N^{+}/Z$).}
\label{radius_and_zeff}
\end{figure}

\begin{figure*}[!t]








\includegraphics[scale=0.212,bb=0 0 2100 640]{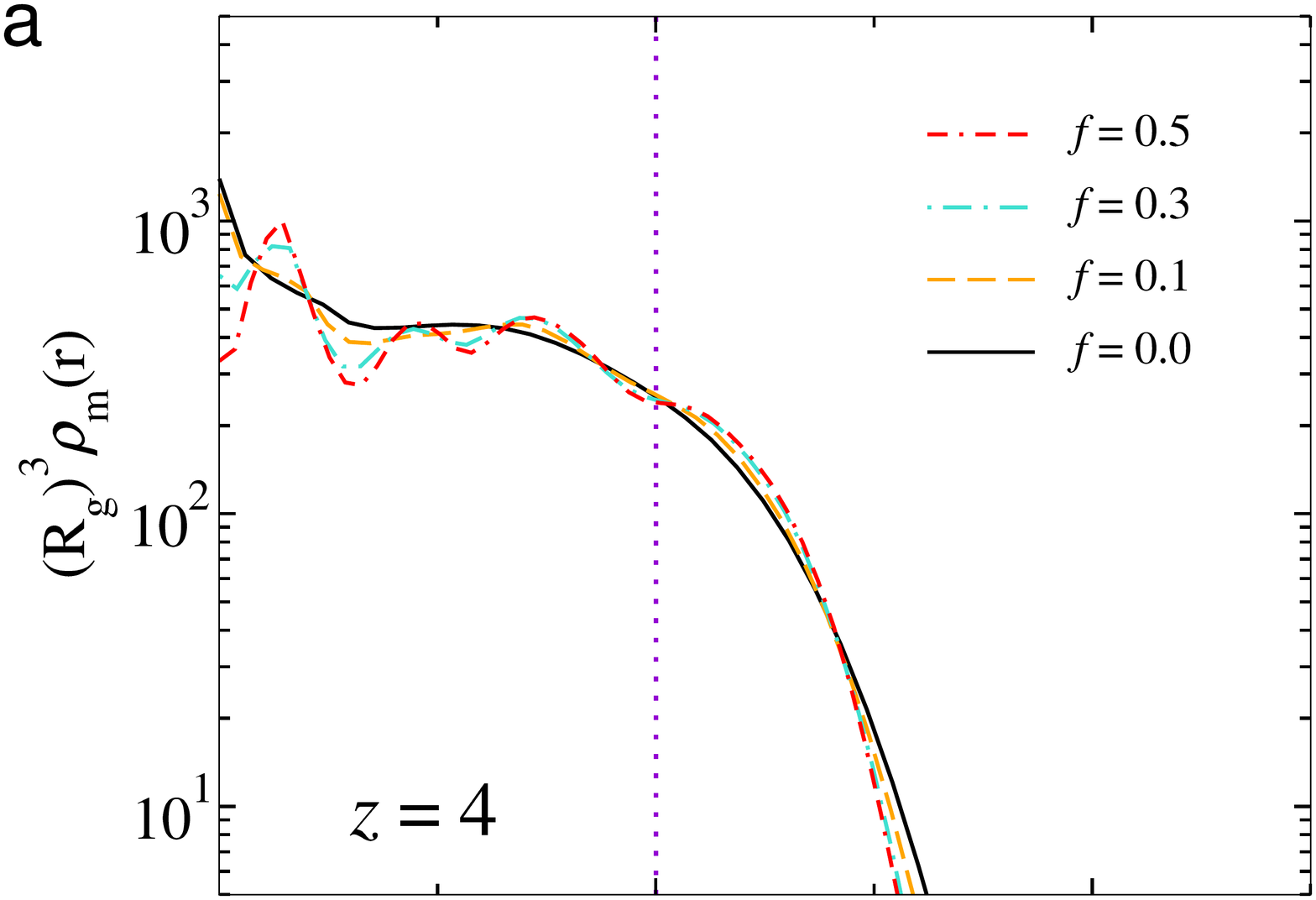}%
\includegraphics[scale=0.212,bb=1300 0 2100 640]{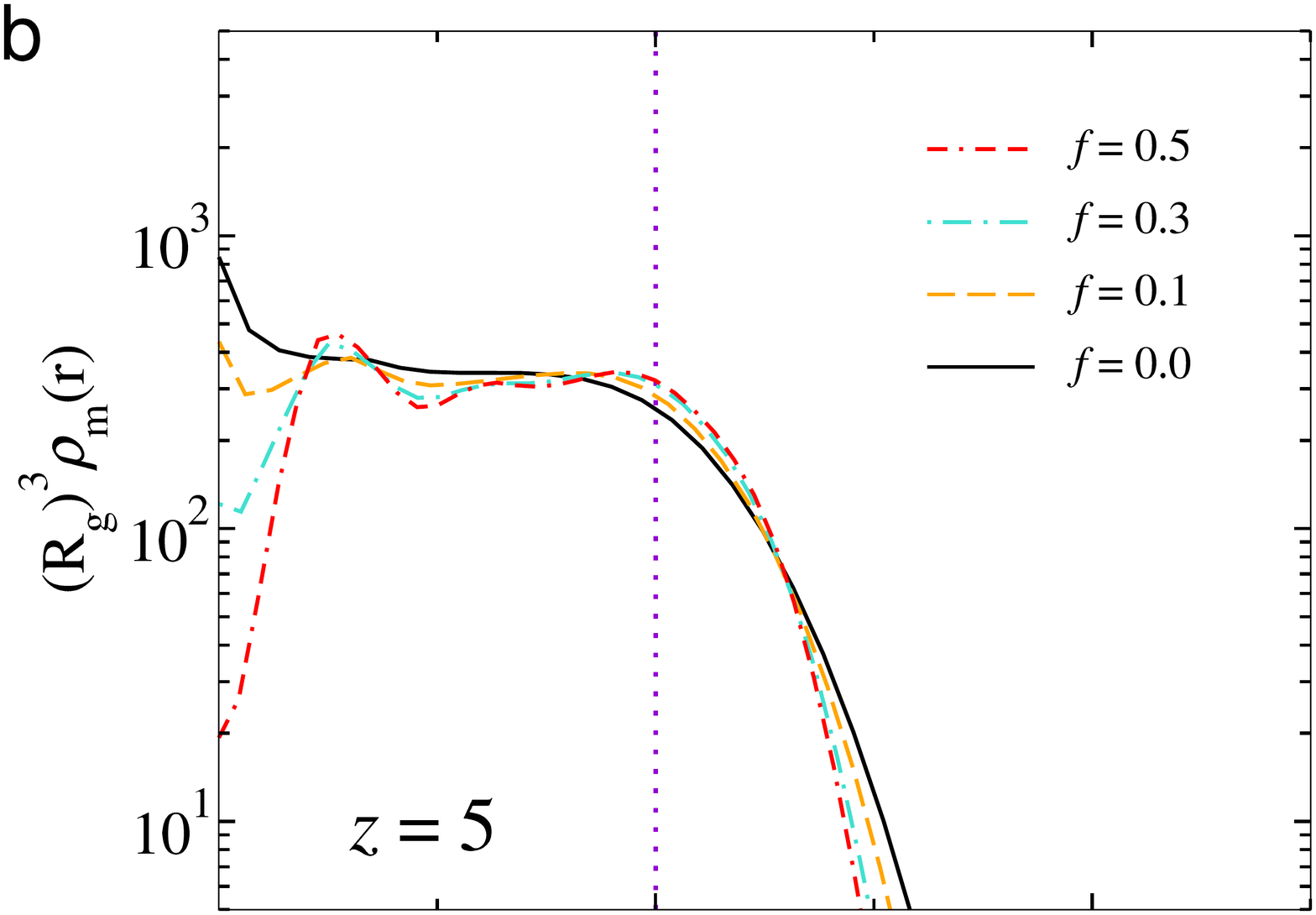}%
\includegraphics[scale=0.212,bb=1300 0 2100 640]{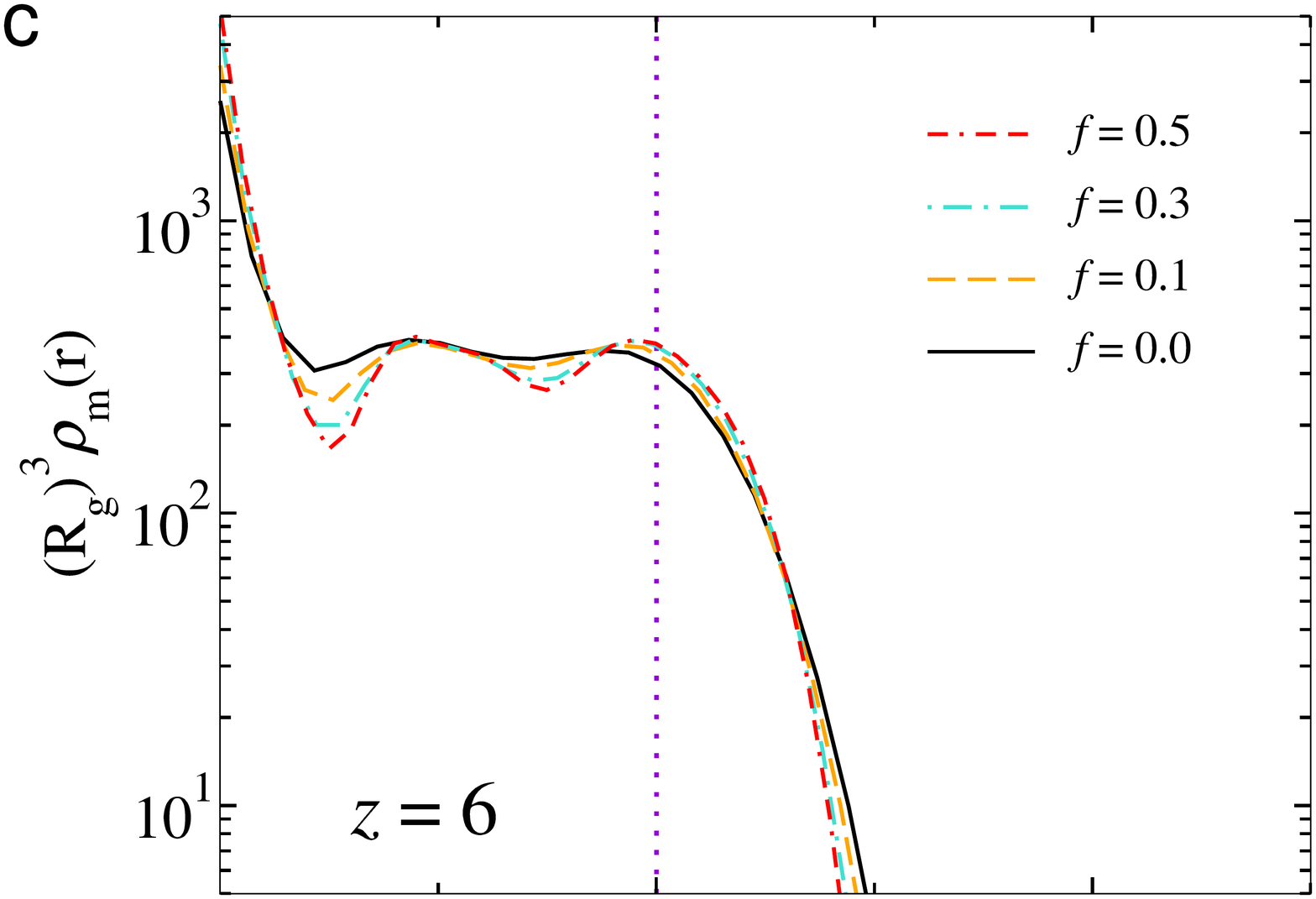}

\vspace{-0.8cm}

\includegraphics[scale=0.212,bb=0 0 2500 640]{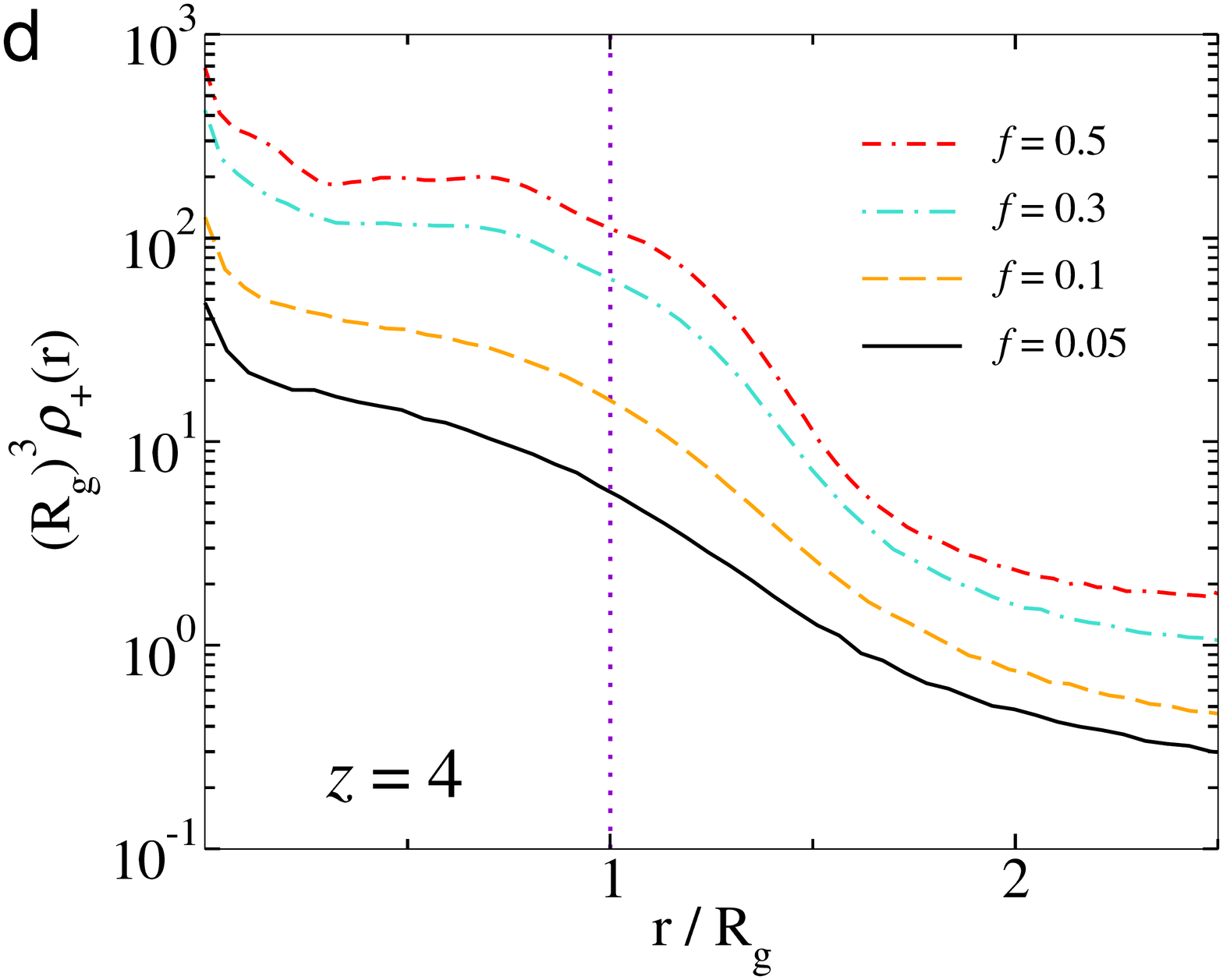}%
\includegraphics[scale=0.212,bb=1700 0 2500 640]{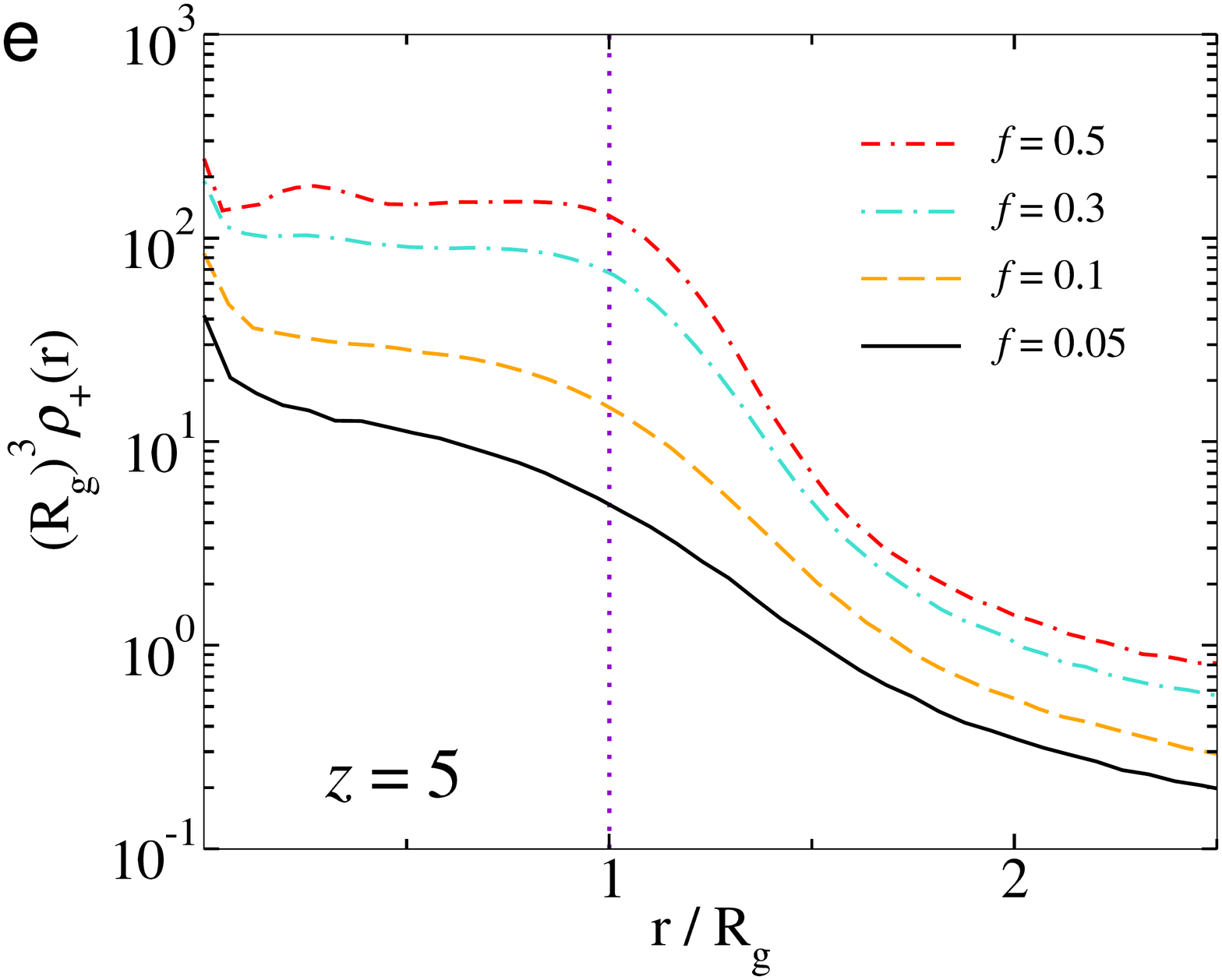}%
\includegraphics[scale=0.212,bb=1700 0 2500 640]{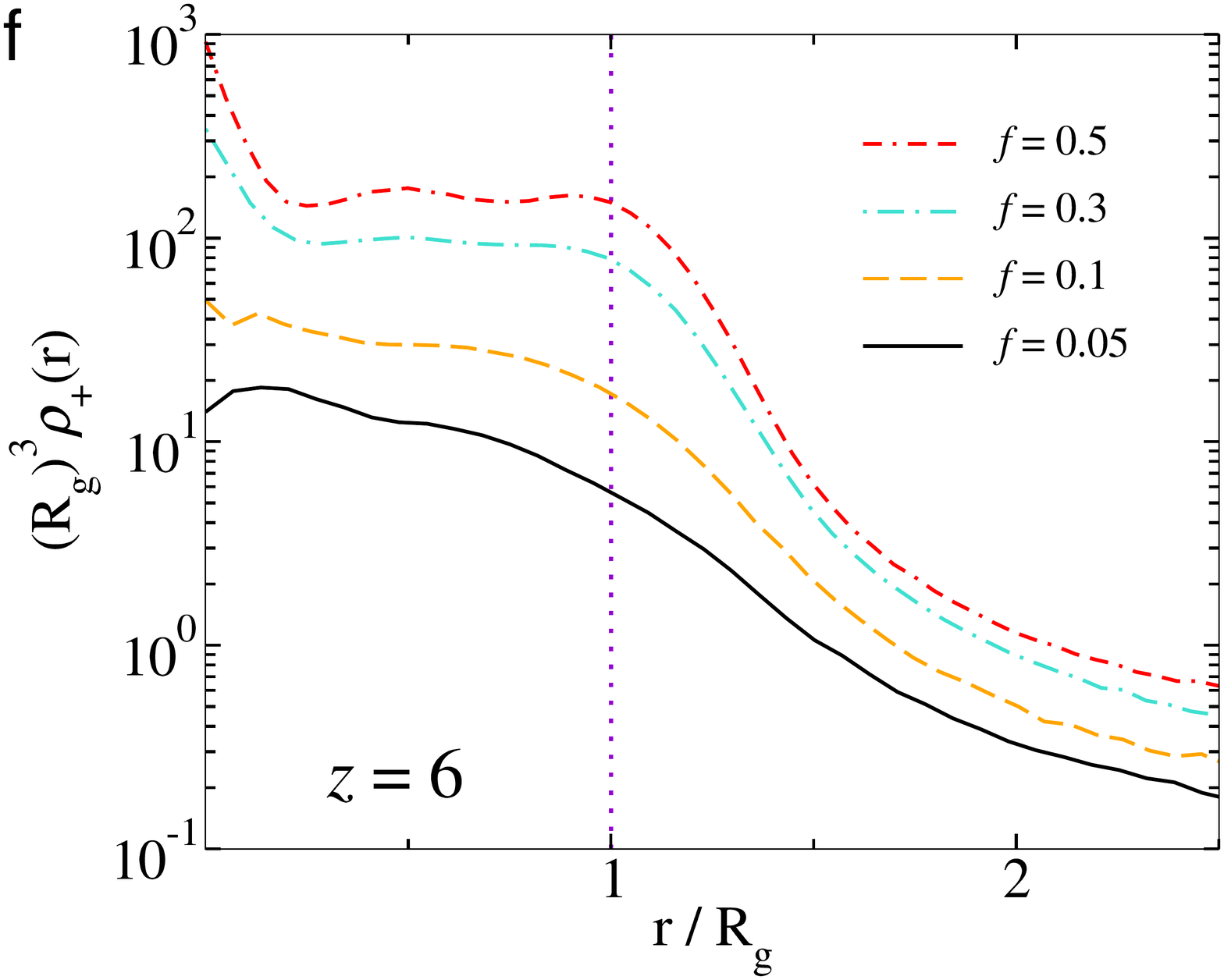}

\caption{Monomer (upper panels) and ionic (bottom panels) density profiles at different values of the dissociation fraction $f$ for different functionality: (a)-(d) $z=4$, (b)-(e) $z=5$, and (c)-(f) $z=6$.}
\label{density_profiles}
\end{figure*}

	First, we investigate the effects of dissociation fraction $f$ on the size of a nanogel particle, which is defined in terms of the radius of gyration $R_{\text{g}}=\left\langle r_{ij}^2 \right\rangle /2 $, where the average $\langle ... \rangle$ is over all $ij$ pairs in the polylectrolyte network~\cite{flory1953book}.
	For instance, when the dissociation fraction is zero, the radius of gyration gives $R_g^{0} =(67\pm 2)$~\AA~for $z=4$, $R_g^{0} =(58\pm 2)$~\AA~ for $z=5$, and $R_g^{0} =(56\pm 2)$~\AA~ for $z=6$.
	We note that these values yields similar results for the volume cross-link densities, $\theta_c = N_c/V_g^0$ with $V_g^0 = 4 \pi (R_{g}^{0})^3/3$, for all networks considered, which contrasts with the line densities $\rho_c$ presented in Table~\ref{parameters}.

	Figure~\ref{radius_and_zeff}(a) shows that the relative nanogel radius $R_g/R_{g}^0$ increases as the dissociation fraction $f$ increases, which is in agreement with what is theoretically predicted in Refs.~\cite{jha2011softmatter,colla2014jcp}.
	The increasing behavior of $R_g$ is similar for the different networks, but one can observe that it gets more pronounced for those with low connectivity.
	Another useful measurement to quantify this effect is the swelling ratio~\cite{microgel2011book}, which can be defined as $S=(V_s - V_d)/ V_d$), where $V_s$ and $V_d$ are the volume of the nanogel in two solvency conditions ({\it e.g.}~the swelled and de-swelled states, respectively).
	Hence for a given dissociation fraction $f$ the swelling ratio can be simply written as $S = ( R_{g}^{3} - (R_{g}^{0})^{3} )/ (R_{g}^{0})^{3}$, with $R_{g}^{0}$ being the nanogel radius for $f=0$.
	As is shown in the inset of Fig.~\ref{radius_and_zeff}(b), the highest swelling ratios $S$ are achieved by the networks formed by cross-links of low connectivity ($z=4$).

	The dissociated counterions can diffuse around the polyelectrolyte network, so that the nanogel particles acquire  an effective net charge~\cite{levin2002pre,claudio2009jcp,chepelianskii2009jphyschemB,chepelianskii2011europhyslett,colla2014jcp}, which is smaller than the bare charge $Z$ of the polyelectrolyte backbone. The effective charge is given by, $Z_{\text{eff}}=Z - (N^{+} - N^{-})$, where $N^{+}$ and $N^{-}$ are the number of positive and negative ions inside the nanogel volume, respectively.
	Here we take $N^{-}=0$ because there are no negative ions in the solution when salt concentration is zero ($C_s=0$).
	Figure~\ref{radius_and_zeff}(b) shows that $Z_{\text{eff}}$ increases with the dissociation fraction $f$, which is consistent with the theoretical predictions of Ref.~\cite{colla2014jcp}.
	For larger values of $f$ the network with functionality $z=4$ presents slightly lower values of $Z_{\text{eff}}$ in comparison to the other networks.
	This occurs mainly because the network with functionality $z=4$ allows for more counterions to become trapped between the polyelectrolytes chains, as shown by the relative number of positive charges inside a nanogel $n^{+}=N^{+}/Z$ (inset of Fig.~\ref{radius_and_zeff}(b)).
	These results corroborate the equilibrium behaviour observed in previous simulations~\cite{claudio2009jcp,quesadaperez2014jcp}.
	However, in contrast to the previous theoretical studies~\cite{levin2002pre,colla2014jcp} that predicted a scaling $Z_{\text{eff}} \sim f^{0.5}$, our results suggest a more linear behavior, which might be related to the relatively small size of the polyelectrolyte networks considered in our simulations.

\begin{figure}[!t]
\centering
\includegraphics[scale=0.32,bb=0 0 750 340]{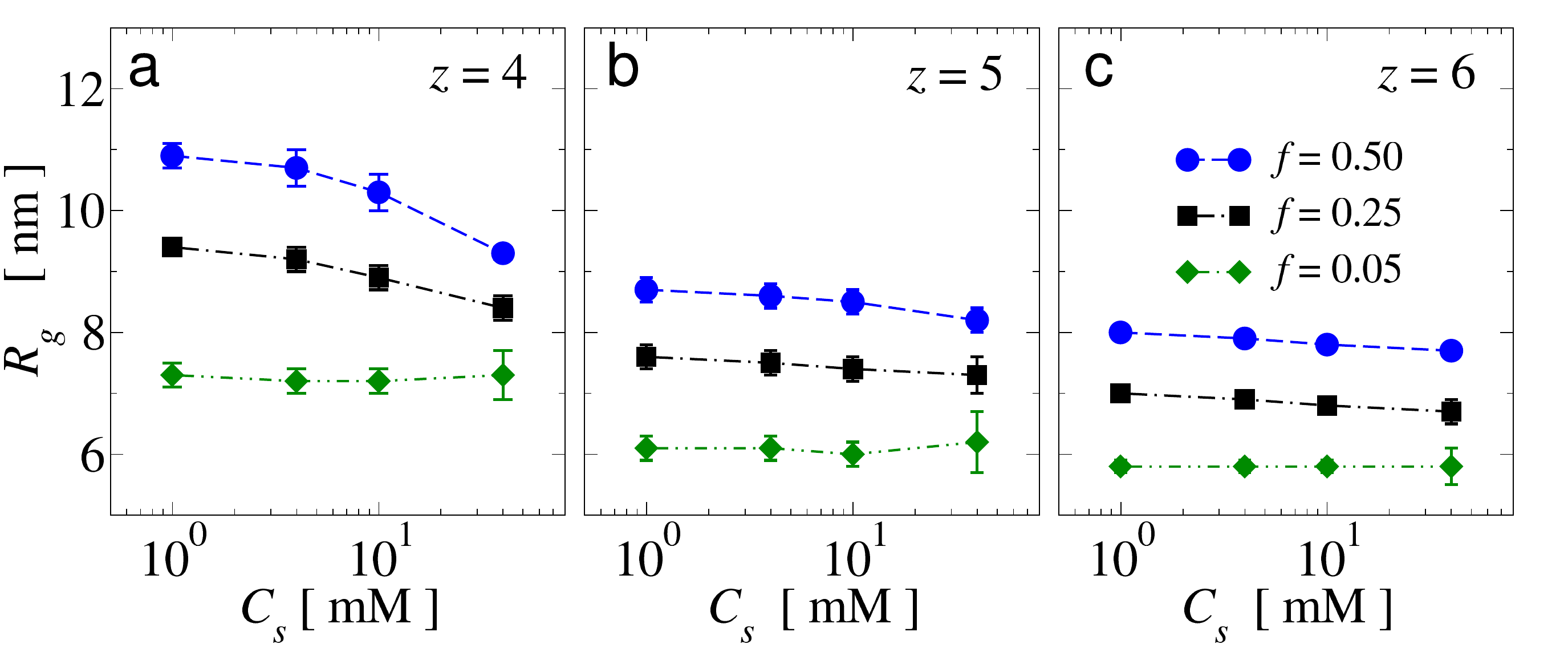}

\caption{Effect of monovalent salt concentration $C_s$ on the size of nanogel particles formed by networks with cross-links of functionality $z$ for different values of the dissociation fraction $f$. (a) $z=4$, (b) $z=5$, and (c) $z=6$.}
\label{salteffect}
\end{figure}


	To better characterize this picture we evaluate the monomer $\rho_{m}(r)$ and the ionic $\rho_{\pm}(r)$ density profiles.
	Figure~\ref{density_profiles} shows $\rho_{m}(r)$ and $\rho_{+}(r)$ for three different networks at different values of dissociation fraction $f$.
	As can be seen in the upper panels, the heterogeneity due to the specific topology of the networks is reflected in monomer density profiles and becomes evident as the dissociation fraction $f$ increases and the nanogel swells.
	Such heterogeneities in density profiles have been observed in simulations and were related both to solvent quality conditions~\cite{kobayashi2014polymers} and to monomer-ion steric repulsion~\cite{adroher2015macromol}.
	Interestingly, it has been speculated that the void created at the center of the nanogel particle might have potential to act as a drug carrier~\cite{quesadaperez2013softmatter}, but our results suggest that the density profiles are inherent to the structure of the network considered, corroborating recent experimental evidences~\cite{dubbert2014macromol}.
	For instance, the monomer density profile in Fig.~\ref{density_profiles}(c) of the swelled network for $f=0.5$ displays a peak when $r/R_g$ approaches the center of the nanogel ($r=0$).
	This happens because a cross-link with functionality $z=6$ is placed (by construction) in the center of that network, which differs from the network with $z=5$ in Fig.~\ref{density_profiles}(b), where no cross-link is placed at the center of the initially generated network.
	Despite these heterogeneities in $\rho_{m}(r)$, the ionic density profiles for the three networks present similar behaviors, {\it i.e.}~showing higher values of $\rho_{+}(r)$ (both inside and outside the nanogel particle) as the fraction $f$ increases.

	Now we turn our attention to the effect of 1:1 salt on the swelling properties of nanogel particles.
	In this case, additionally to the $Z$ positively charged ions due to dissociation of the ionizable groups of the polyelectrolyte network, the dissociation of salt at a given concentration $C_s = N_s/ V$ leads to $N_s$ coions (anions) and $N_s$ counterions (cations) in solution.
	For simplicity, such additional ions are considered to be hard spheres which interact with all charged particles in the system via Coulomb (Eq.~\ref{coulombpot}) and hard core potentials.

	As shown in Fig.~\ref{salteffect}, the effect of the presence of salt on the de-swelling behavior of the network is weaker for the networks with high connectivity.
	Particularly, for the network with $z=4$, one can see that while the nanogel radius $R_g$ remains constant for low values of the dissociation fraction, it displays a decreasing behavior for $f=0.5$, which is in agreement with the behavior predicted by the theory of Ref.~\cite{colla2014jcp}, and also observed in 
experiments~\cite{shibayama1996jcp,nisato1999langmuir,fernandez2001jcp,lopezleon2006jphyschemB,capriles2008jphyschemB,barbero2009review}and in a previous simulation study~\cite{quesadaperez2014jcp}.
	We note that even though excluded volume effects are taken into account in our model, we are not able to observe a reentrant (or re-swelling) behavior expected for high salt concentrations~\cite{jha2012softmatter,sing2013macromol}.



\section{CONCLUDING REMARKS}

	In summary, we presented an analysis of the swelling behavior of ionic nanogels defined by explicit polyelectrolyte networks formed by cross-links of different topologies.
	Our results indicate that despite the higher effective cross-link line density $\rho_c$, the polyelectrolyte network formed by cross-links with low connectivity ($z=4$) display the highest volume changes both in the presence and in the absence of monovalent salt in the system.
	In particular, we have verified that both nanogel size $R_g$ and its effective net charge $Z_{\text{eff}}$ increase as the fraction $f$ of ionizable groups in the network increases.
	Also, there is a clear de-swelling behavior for a increasing concentration of salt in solution for the networks with functionality $z=4$ at high values of $f$.
	We found that these effects are dramatically reduced for the networks with functionality $z=6$, indicating that a higher elastic energy contribution is hampering the change in the size of the nanogel particle, which is confirmed in the Appendix 
	section.~By considering that the effective number of chains in the network $N_{\text{eff}}$ is proportional to the functionality $z$ of its cross-links, our results corroborates Flory's assumption that $\Delta F_{\text{elastic}} \propto N_{\text{eff}}$, although in his theory the elastic free energy has a purely entropic origin~\cite{flory1985britpolj}.

	Finally, we note that our results are in a qualitative agreement with the swelling behavior theoretically  predicted in Ref.~\cite{colla2014jcp}.
	A quantitative comparison with the theory may be hampered by the assumptions of implicit chain (modeled by an elastic energy contribution) and a flat monomer density inside the nanogel volume, as these approximations might  not be appropriate for small-sized nanogel particles studied in the present simulations.


\section*{Acknowledgement}

	The authors acknowledge Alexandre P. dos Santos and Thiago E. Colla for the useful discussions, and the financial support from CNPq (Grant No~165153/2014-8), INCT-FCx, and by the US-AFOSR under the grant FA9550-12-1-0438.


\section*{Appendix: Elastic energy of the polyelectrolyte networks}

\begin{figure}[!h]
\centering

\includegraphics[scale=0.31,bb=0 0 750 650]{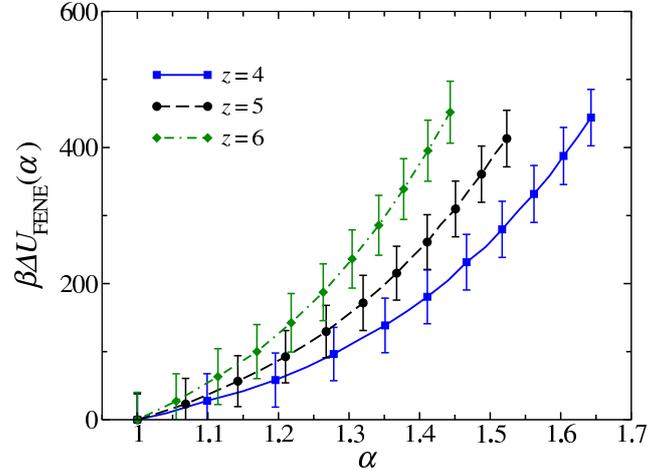}

\vspace{-0.5cm}

\caption{Change in the elastic energy $\beta \Delta U_{\text{FENE}}$ as function of the ratio $\alpha=R_g/R_g^0$ for nanogel particles with cross-links of functionality $z$. (lines are guides to the eyes only).}
\label{comparisonFLORY}
\end{figure}

	Here we present an analysis of the elastic energy obtained from simulations of a salt-free solution  where different values of the dissociation fraction $f$ yields networks with different sizes, which are described by the ratio $\alpha=R_g/R_g^0$ (see Fig.~\ref{radius_and_zeff}a).
	Estimates for the changes in elastic energy are obtained as the difference in the equilibrium values of the FENE potential (see Eq.~\ref{fenepotential} for the pairwise definition) evaluated for the whole network at two dissociation fraction conditions, {\it i.e.}~$\langle \beta \Delta U_{\text{FENE}} \rangle (f) = \langle \beta U_{\text{FENE}} \rangle (f) - \langle \beta U_{\text{FENE}} \rangle (0)$.
	As shown in Figure~\ref{comparisonFLORY}, the elastic energy contribution is higher for the networks with high connectivity.

%

\end{document}